\documentclass{amsart}
\usepackage{amsmath}

 \usepackage{graphics}
\usepackage{graphicx}

\usepackage{amssymb}
 \usepackage{amsthm}

  \usepackage{lineno}

\newtheorem{theorem}{Theorem}[section]
\newtheorem{lemma}{Lemma}[section]

\newtheorem{corollary}{Corollary}[section]
\newtheorem{definition}{Definition}[section]
 
\newtheorem{example}{Example}[section]
\newtheorem{remark}{Remark}[section]

\begin{document}

\title{Finite symmetric   functions with non-trivial arity gap}

\author{Slavcho Shtrakov \and J\"org  Koppitz}

\address{Department of Computer Science, South-West University,  2700
Blagoevgrad, Bulgaria} \email{shtrakov@swu.bg} \urladdr{http://shtrakov.swu.bg}
\address{
Institute of Mathematics, University of
Potsdam, 14415 Potsdam, Germany
}\email{koppitz@uni-potsdam.de}
\keywords{symmetric function, essential variable, subfunction, identification
minor, essential arity
gap, gap index, separable set.
}

 \subjclass[2000]{Primary: 94C10; Secondary: 06E30\\
 ~~~~{\it ACM-Computing Classification System (1998)} : G.2.0}

\begin{abstract}
Given an $n$-ary
  $k-$valued function  $f$, $gap(f)$ denotes the essential arity gap of $f$ which is the minimal number of essential variables  in $f$ which become fictive when identifying any two distinct essential variables in $f$.
In the present paper we study the properties of  the symmetric function with non-trivial arity gap ($2\leq gap(f)$).
We prove several results concerning decomposition of the symmetric functions with non-trivial arity gap with its minors or subfunctions.  We show that   all non-empty sets of essential variables in   symmetric functions with non-trivial arity gap  are separable. 
\end{abstract}
\maketitle



\section*{Introduction}~~~

Given a function $f$, the essential variables in $f$ are defined as variables
which occur in $f$ and weigh with the values of that function. They are
investigated when replacing the variables with constants or variables (see e.g.
\cite{ch3,ch51,sal,sh51}).  If we replace in a function $f$ some  variables with
constants the result is a subfunction of $f$ and when replacing several  variables
with other variables, the result is a minor of $f$.
 
 The essential arity gap of a finite-valued function $f$  is the minimum
decrease in the number of essential variables in  identification minors of $f$.
In this paper we investigate functions in $k$-valued logics with non-trivial arity gap, which are
important in theoretical and applied computer sciences, namely the
symmetric functions. 
 
 R. Willard proved that if a function $f: A^n\rightarrow B$ depends on $n$
variables and  $k<n$, where $k=|A|$ then $gap(f)\leq 2$   \cite{ros}. On the
other side it is clear that $gap(f)\leq n$. Thus in any case we have $gap(f)\leq
min(n,k)$.

M. Couceiro and  E. Lehtonen proposed 
 a classification of functions according to their arity gap \cite{mig1,mig2}.
 
We have  proved that if $2\leq gap(f)<min(n,k)$ then $f$ can be decomposed as a sum of functions of a prescribed type (see Theorem 3.4 \cite{sl2}). 
 
A natural question to ask is which additional properties, concerning arity gap are typical for the symmetric and linear functions with non-trivial arity gap.
We investigate the behavior of the subfunctions of symmetric functions with non-trivial arity gap. So, in the present paper we consider together the both types of replacement in function's inputs -  with constants (subfunctions) and with variables (minors). We  prove that   ``almost'' all  subfunctions of a symmetric function $f$ with non-trivial arity gap inherit the property of $f$ concerning the identification of variables.
 We are interesting also in decomposition of symmetric   functions as ``sums of conjunctions`` (following \cite{sl2}). 

We also characterize the relationship between separable sets and subfunctions of symmetric functions with non-trivial arity gap.

\section{Preliminaries}\label{sec1}

Let $k$ be a natural number
with $k\geq 2$. Denote by $K=\{0,1,\ldots,k-1\}$ the set (ring) of
 remainders modulo $k$. An {\it $n$-ary $k$-valued function
(operation) on $K$ } is a mapping $f: K^n\to K$ for some natural
number $n$, called {\it the arity} of $f$. The set of all $n$-ary $k$-valued
functions is denoted by $P_k^n$.

  Let $f\in P_k^n$ and   $var(f)=\{x_1,\ldots,x_n\}$  be the set of all variables, which occur in $f$.
We say that the $i$-th variable $x_i\in var(f)$ is  {\it  essential} in $f(x_1,\ldots,x_n)$, or $f$ {\it
essentially depends} on $x_i$, if there exist values
$a_1,\ldots,a_n,b\in K$, such that
\[
   f(a_1,\ldots,a_{i-1},a_{i},a_{i+1},\ldots,a_n)\neq
   f(a_1,\ldots,a_{i-1},b,a_{i+1},\ldots,a_n).
\]

The set of all essential variables in the function $f$ is denoted by
$Ess(f)$ and the number of its essential variables  is denoted by
$ess(f):=|Ess(f)|$.

Let $x_i$ and  $x_j$ be two distinct essential variables in $f$. The
function $h$ is obtained from $f\in P_{k}^{n}$ by {\it the
identification of the variable $x_i$ with $x_j$}, if
\[
  h(a_1,\ldots,a_{i-1},a_i,a_{i+1},\ldots,a_n):=f(a_1,\ldots,a_{i-1},a_j,a_{i+1},\ldots,a_n),
\]
for all $(a_1,\ldots,a_n)\in K^n$.

Briefly, when  $h$ is obtained from $f,$ by identification of the
variable $x_i$ with $x_j$, we will write $h=f_{i\leftarrow j}$ and $h$ is
called {\it an identification minor of $f$}.   
 Clearly, $ess(f_{i\leftarrow j})\leq ess(f)$, because
$x_i\notin Ess(f_{i\leftarrow j})$, even though it might be essential in
$f$.
When $h$ is an identification minor of $f$ we shall write $f\vdash h$. The transitive closure of $\vdash$ is denoted by $\models$.  $Min(f)=\{h\ | \ f\models h\}$ is the set of all minors of $f$.

Let  $f\in P_k^n$ be an $n$-ary $k$-valued function. Then   {\it the essential arity gap} (shortly
{\it arity gap} or {\it  gap}) of $f$ is defined by
\[gap(f):=ess(f)-\max_{h\in Min(f)}ess(h).\]

Let $h\in Min(f)$ be a minor of $f$ and 
\[L_h:=\{m\ |\ \exists\ (h_1,\ldots,h_m)\ \mbox{ with } f\vdash h_1\vdash\ldots\vdash h_m=h\}.\] The number {$depth(h):=\max L_h$} is called \emph{the depth} of $h$ and \emph{the gap index} of $f$ is defined as follows
\[ind(f):=\max_{h\in Min(f)}depth(h).\]

Let $2\leq p\leq m$. We let $G_{p,k}^m$ denote the set of all $k$-valued functions
which essentially  depend on $m$ variables whose arity gap is equal to
$p$, i.e. $G_{p,k}^m=\{f\in P_k^n\ |\ ess(f)=m\ \&\ gap(f)=p\}$.
 
Let $x_i$  be an essential variable in $f$ and $c\in K$ be a constant from $K$. The
function $g:=f(x_i=c)$ obtained from $f\in P_{k}^{n}$ by replacing the variable $x_i$ with $c$  is
called a {\it simple subfunction of $f$}.   
 
When $g$ is a simple subfunction of $f$ we shall write $f\rhd g$. The transitive closure of $\rhd$ is denoted by $\gg$.  $Sub(f)=\{g\ | \ f\gg g\}$ is the set of all subfunctions of $f$ and $sub(f):=|Sub(f)|$.

Let $g\in Sub(f)$ be a subfunction of $f$ and let 
\[O_g:=\{m\ |\ \exists\ (g_1,\ldots,g_m)\ \mbox{ with } f\rhd g_1\rhd\ldots\rhd g_m=g\}.\] The number $ord(g):=\max O_g$ is called \emph{the order} of $g$.

As usual we  denote by $S_n$ the set of all permutations of the set
$\{1,\ldots,n\}$.
Let
$Ess(f)=\{x_{i_1},\ldots,x_{i_m}\}\subseteq \{x_1,\ldots,x_n\}$.    Let $S_f$ be
the set of all  permutations of $\{{i_1},\ldots,{i_m}\}$. We say that $f$ is a
\emph{symmetric function}  if
 $f(x_1,\ldots,x_n)=f(x_{\pi(1)},\ldots, x_{\pi(n)}),$ 
for all $\pi\in S_f$.

Given a variable $x$ and $c\in K$, $x^c$ is an unary
 function defined by:
\[
  x^c=\left\{\begin{array}{ccc}
             1 \  &\  if \  &\  x=c \\
             0 & if & x\neq c.
           \end{array}
           \right.
\]

 We  use {\it sums of
conjunctions (SC)} for representation of functions in $P_k^n$. This is the most natural representation of the functions in finite algebras. It is based on so called operation tables of the functions.

Each function $f\in P_k^n$ can be uniquely represented in SC-form as
follows
\[
    f=a_0\cdot x_1^{0}\ldots x_n^{0}\oplus\ldots\oplus
     a_{m}\cdot x_1^{c_1}\ldots
     x_n^{c_n}\oplus\ldots\oplus a_{k^n-1}\cdot x_1^{k-1}\ldots
     x_n^{k-1}
\]
with  $m=\Sigma_{i=1}^{n} c_i k^{n-i}$, and $c_i, a_{m}\in K$, where $"\oplus"$ and
$"\cdot"$ are the operations addition and multiplication modulo $k$ in
the ring $K$.

\section{Symmetric  functions with non-trivial arity gap}\label{sec2}

We are going to  study the behavior of the symmetric $k$-valued functions $f$
with non-trivial arity gap, i.e. with  $gap(f)>1$.

\begin{lemma}\label{l2.1}
Let $f\in P_k^n$ be a symmetric function which essentially depends on $n$ variables and let $f\gg g$ then $g$ is a symmetric function and if $Ess(g)\neq\emptyset$ then $ess(g)= n-ord(g)$. 
\end{lemma}
\begin{proof}
Without loss of generality let us assume 
  that  $ord(g)=m>0$ and \[f\rhd f(x_1=c_1)\rhd\ldots\rhd f(x_1=c_1,x_2=c_2,\ldots,x_m=c_m)=g.\]
It is obvious that $g$ is symmetric.

Clearly,   $x_{i}\in Ess(g)$ if and only if  $x_j\in Ess(g)$ for all $i,j\in \{ m+1,\ldots,n\}$. Hence if $Ess(g)\neq\emptyset$ then $Ess(g)=X_n\setminus \{x_1,\ldots,x_m\}$.
\end{proof}

\begin{lemma}\label{l2.2} Let  $2\leq p\leq min(k,n)$.
If $f\in G_{p,k}^n$ is a symmetric function, then $p=2$ or $p=n$.
\end{lemma}
\begin{proof}
 Let us suppose this is  not the case. Then $2<p<n$. Hence there is an
identification minor $h$ of $f$ such that $gap(f)=n-ess(h)$ and $2<n-ess(h)<n$.
Without loss of  generality assume that $h=f_{n\leftarrow n-1}$ and
$Ess(h)=\{x_1,\ldots,x_q\}$, where $q=n-p$ such that $0<q< n-2$. Then
$x_{n-2}\in Ess(f)\setminus Ess(h)$. Hence for every $n$ constants
$c_1,\ldots,c_{n-3},c_{n-2},d_{n-2},c_{n-1}\in K$ we have
\\ \centerline{$f(c_1,\ldots,c_{n-3},c_{n-2},c_{n-1},c_{n-1})=f(c_1,\ldots,c_{n-3},
d_{n-2},c_{n-1},c_{n-1}).$}
Since $f$ is symmetric, Lemma \ref{l2.1} implies  \\ \centerline{
$f(c_{n-2},\ldots,c_2,c_{1},x_{n-1},x_{n-1})=f(d_{n-2},c_{n-3},\ldots,c_2,c_{1},
x_{n-1},
x_{n-1})$.}
 Hence $x_1\notin Ess(h)$, which is a contradiction.
\end{proof} ~~~

\begin{lemma}~\label{l2.3}   \cite{sl2}
Let $f$ be a $k$-valued  function which depends essentially on all of its $n$, $n>3$ variables and $gap(f)=2$. Then there exist   two distinct essential variables $x_u,x_v$ such that  $ess(f_{u\leftarrow v})=n-2$ and $x_v\notin Ess(f_{u\leftarrow v})$.
Moreover, $ess(f_{u\leftarrow m})=ess(f_{v\leftarrow m})=n-2$ for all $m$, $1\leq m\leq n$ with $m\notin \{u,v\}$.
\end{lemma}

\begin{lemma}\label{l2.4} Let $3<n\leq k$. 
If $f\in G_{2,k}^n$ is a symmetric function then 
 $x_v\notin Ess(f_{u\leftarrow v})$ for all $1\leq
u,v\leq n$ with $u\neq v$.
\end{lemma}
\begin{proof}
 From Lemma \ref{l2.3}, there are  $1\leq
u,v\leq n$ with $u\neq v$ such that $x_v\notin Ess(f_{u\leftarrow v})$. Without loss of generality, let $u=1$ and $v=2$. Further, let $1\leq i<j\leq n$ and $a_{1},\ldots ,a_{n},b\in K$. Then we have \[f_{i\leftarrow j}(a_{1},\ldots ,a_{n})= f(a_{1},\ldots
,a_{i-1},a_{j},a_{i+1},\ldots ,a_{n})=\] \[f(a_{j},a_{j},a_{1},\ldots
,a_{i-1},a_{i+1},\ldots ,a_{j-1},a_{j+1},\ldots ,a_{n})= \]
\[f(b,b,a_{1},\ldots ,a_{i-1},a_{i+1},\ldots ,a_{j-1},a_{j+1},\ldots
,a_{n})=\] \[f(a_{1},\ldots ,a_{i-1},b,a_{i+1},\ldots ,a_{j-1},b,a_{j+1},\ldots
,a_{n})=\] \[f_{i\leftarrow j}(a_{1},\ldots ,a_{j-1},b,a_{j+1},\ldots ,a_{n}).\]
This shows that $x_{j}\notin Ess(f_{i\leftarrow j})$.
\end{proof}
 
\begin{remark}\label{r2.1}
 If $f$ is a symmetric function with non-trivial arity gap then all its identification minors are symmetric. In fact, we have  $h=f_{2\leftarrow 1}=f(c,c,x_3,\ldots,x_n)$ for all $c\in K$, according to Lemma \ref{l2.4}. Hence $h$ is the subfunction $h=f(x_1=c,x_2=c)$ of $f$ and by Lemma \ref{l2.1}  it follows that $h$ is symmetric.
\end{remark}

\begin{lemma}\label{l2.5}
If $f\in G_{2,k}^n$, $n\geq 2$,  is a symmetric function then $1\leq ind(f)\leq \frac{n}{2}$. 
\end{lemma}
\begin{proof}
 Clearly if $ess(f)\geq 2$ then $ind(f)\geq 1$ for all $f\in P_k^n$.

Lemma \ref{l2.3} and  Lemma \ref{l2.4} imply that if $f\vdash
h_1\vdash\ldots\vdash h_m$ with $m=ind(f)$ then $depth(h_i)=i$ and 
$ess(h_i)=n-2i$ for $i=1,\ldots,m$. Hence $ind(f)\leq \frac{n}{2}$.
\end{proof}

Let $f\in G_{2,k}^n$, $n>2$,  be a symmetric function and let $ind(f)=m<\frac{n}{2}$. Then for each minor $h\in Min(f)$ with $depth(h)<m$ there is $g\in Min(f)$ such that $f\models h\models g$ and $depth(g)=m$. 
 
\begin{remark}\label{r2.2}
 Let $f\in G_{2,k}^n$, $n>2$,  be a symmetric function and let $h\in Min(f)$. From
Lemma \ref{l2.2}, we conclude that  if $depth(h)=l<ind(f)$, then $h\in G_{2,
k}^{n-2l}$, else $h\in G_{n-2l, k}^{n-2l}$.
\end{remark}

Let $ k$ and $n$, $k\geq n>1$, be two natural numbers such that $1< n\leq k$.
The set $K^n$ of all $n-$tuples over $K$ is the disjoint union of the
following two sets:
\\ \centerline{$Eq_k^n:=\{(c_1,\ldots,c_n)\in K^n\  |\ \
c_i=c_j,
\mbox{ for some } i,j \mbox{ with }\ i\neq j\},$} 
\\ \centerline{$Dis_k^n:=\{(c_1,\ldots,c_n)\in K^n\  |\ \
c_i\neq c_j,
\mbox{ for all } i,j \mbox{ with }\ i\neq j\}. $}

\begin{theorem}\label{t2.1} \cite{sl2}  Let $3\leq n\leq k$. Then  $f\in
G_{n,k}^n$, if and only if $f$ can be  represented as follows
\begin{equation}\label{eq2.1}~~
f=[\bigoplus_{{\beta}\in Dis_k^n}
a_{{\beta}}.x_1^{d_1}\ldots x_n^{d_n}]\ \oplus\
a_0.[\bigoplus_{\alpha\in
Eq_k^n}x_1^{c_1}\ldots x_n^{c_n}],\
 \end{equation}
  where   $\beta=(d_1,\ldots,d_n)$ and
$\alpha=(c_1,\ldots,c_n)$, and
at least two
among  the coefficients $a_0,a_{{\beta}}\in K$ for ${\beta}\in Dis_k^n$,
 are distinct. 
\end{theorem}

Let $\alpha=(c_1,\ldots,c_n)\in K^n$. We denote
\[S(n,{\alpha}):=\bigoplus_{\pi\in S_n}x_1^{c_{\pi(1)}}\ldots
x_n^{c_{\pi(n)}}.\]

Let $\alpha=(c_1,\ldots,c_n)\in K^n$ and $\beta=(d_1,\ldots,d_m)\in K^m$ with $m\leq n$.
 
 We shall write $\beta\leq \alpha$ if there are $1\leq
i_1,\ldots,i_m\leq n$ such that $c_{i_j}=d_j$  and $c_s\neq
d_j$ for all $s\notin \{i_1,\ldots,i_m\}$ and all $j\in \{1,\ldots, m\}$. 

\begin{example}\label{ex2.1} Let $k=5$. Then  $(0,1,1)\leq (0,1,2,1,4)$, but
$(0,1)\not\leq (0,1,2,1,4)$ and $(0,2,3)\not\leq (0,1,2,1,4)$. Let
$\alpha=(1,2,4)$. Then 
\[S(3,{\alpha})=x_1^1x_2^2x_3^4\oplus x_1^1x_2^4x_3^2\oplus
x_1^2x_2^1x_3^4\oplus x_1^2x_2^4x_3^1\oplus x_1^4x_2^1x_3^2\oplus
x_1^4x_2^2x_3^1.\] 

\end{example}

\begin{theorem}\label{t2.2}
Let $f\in G_{n,k}^n$, $3\leq n\leq k$. Then $f$ is a  symmetric function if and only if it can be represented in the following form:
\begin{equation}\label{eq2.2}
f=a_0\big[
\bigoplus_{{\alpha}\in Eq_k^n}x_1^{c_1}x_2^{c_2}\ldots
x_n^{c_n} \big]\oplus \big[\bigoplus_{\beta\in Dis_k^n}b_{\beta}S(n,{\beta})\big],\end{equation}
where  ${\alpha}=(c_1,c_2,\ldots,c_n)\in Eq_k^n$, and at
least two
among  the coefficients $a_0,b_{\beta}\in K$, for ${\beta}\in Dis_k^n$
 are distinct.
                                                                                
 \end{theorem}

\begin{proof}
        Let   $f\in G_{n,k}^n$, $2<n\leq k$ be a symmetric function and  
${\beta}=(d_1,\ldots,d_n)\in Dis_k^n$. Let us put
$b_{{\beta}}=f({\beta})$. Since $f$ is a symmetric function, it follows that 
 $f(d_{\pi(1)},d_{\pi(2)},\ldots,d_{\pi(n)})=b_{{
\beta}},$
for each $\pi\in S_n$.

Let $\alpha\in Eq_k^n$. Then (\ref{eq2.1}) implies
$f(\alpha)=f(0,0,\ldots,0)=a_0$, which proves that $f$ is represented in
the form (\ref{eq2.2}).
Clearly, if $f$ is represented as in (\ref{eq2.2}), then it is symmetric function, whose arity gap is equal to $n$.
\end{proof}
\begin{corollary}\label{c2.1}
There are $k^{\binom{k}{n}+1}-k$   different symmetric
functions in $G_{n,k}^n$.
\end{corollary}
\begin{proof}
 There exists $\binom{k}{n}$ ways to choice $\beta$ in
(\ref{eq2.2}). Thus there are $\binom{k}{n}+1$ coefficients in (\ref{eq2.2}),
including $a_0$ taken from $K$. On the other hand we have to exclude all $k$ cases when 
$a_0=b_\beta$  for $\beta\in Dis_k^n$. 
\end{proof}
We are interesting in  explicit representation of the symmetric functions $f$ with
$gap(f)=2$ in the case  when $ess(f)=3$. The case $gap(f)=2$ and
$ess(f)=3$ is really special which is deeply discussed in \cite{sl2} where we
decomposed $f\in G_{2,k}^3$ for $k=3$ (see Theorem 5.1 \cite{sl2}). In a
similar way one can prove the following more general result.
\begin{theorem}\label{t2.3}
Let $f\in G_{2,k}^3$, $k\geq 3$. Then $f$ is a  symmetric function if and only if it can be represented in one of the following forms: 
\begin{equation}\label{eq2.3} f=\bigoplus_{i=0}^{k-1} a_i\big[x_1^ix_2^ix_3^i
 \oplus \big[\bigoplus_{\alpha\in Eq_k^3, \ (i)\leq\alpha} x_1^{c_1}x_2^{c_2}x_3^{c_3}\big]\big]
 \oplus \big[\bigoplus_{\delta\in Dis_k^3}b_{\delta}S(3,\delta)\big]\end{equation}
or
\begin{equation}\label{eq2.4} f=\bigoplus_{i=0}^{k-1} a_i\big[x_1^ix_2^ix_3^i
 \oplus \big[\bigoplus_{\alpha\in Eq_k^3, \ (ii)\leq\alpha} x_1^{c_1}x_2^{c_2}x_3^{c_3}\big]\big]
 \oplus \big[\bigoplus_{\delta\in Dis_k^3}b_{\delta}S(3,\delta)\big],\end{equation}
  where $\alpha=(c_1,c_2,c_3)$ and at
least two
among  the coefficients   $a_i\in K$, for    $i=0,\ldots, k-1$
 are distinct.
\end{theorem}

\begin{theorem}\label{t2.4}
Let $f\in P_{k}^n$ be a symmetric function with non-trivial arity gap. Then

 $(i)$ If $gap(f)=n$ or  $n$, $n\geq 2$, is an even natural number or $ind(f)<\frac{n-1}{2}$ then 
$f(c_1,\ldots,c_1)=f(c_2,\ldots,c_2)$ for all $c_1,c_2\in K$;

$(ii)$  If  $n$, $3\leq n\leq k$, is an odd natural number, $gap(f)=2$ and $ind(f)=\frac{n-1}{2}$ then there exist at least two values $c_1,c_2\in K$ such that 
$f(c_1,\ldots,c_1)\neq f(c_2,\ldots,c_2)$. 
\end{theorem}
\begin{proof}
$(i)$ We have to consider three cases:

{\bf Case A.} Let $gap(f)=n$.

Then $f\in G_{n,k}^n$  and from Theorem \ref{t2.1} 
it follows $f(c_1,\ldots,c_1)=
f(c_2,\ldots,c_2)$ for all $c_1,c_2\in K$.

 {\bf Case B.} Let $n$, $n\geq 2$ be an even natural number and $gap(f)=2$.

 Let $c_1,c_2\in K$
be  two constants with
$c_1\neq c_2$.
 From Lemma \ref{l2.4} it follows that $x_v\notin Ess(f_{u\leftarrow v})$ for all $1\leq
u,v\leq n$ with $u\neq v$.
Then we obtain \\
 \begin{tabular}{ll}
  $f(c_2,c_2,\ldots,c_2)$& \\
=$f(c_1,c_1,c_2,c_2,c_2,\ldots,c_2)$&because $x_2\notin Ess(f_{1\leftarrow 2})$\\
=$f(c_1,c_1,c_1,c_1,c_2,\ldots,c_2)$& because $x_3\notin Ess(f_{4\leftarrow 3})$\\
=$f(c_1,c_1,c_1,c_1,c_1,c_1,c_2,\ldots,c_2)$& because $x_5\notin Ess(f_{6\leftarrow 5})$\\
\ldots\quad\ldots & \ldots\quad\ldots \\
=$f(c_1,c_1,\ldots,c_1,c_1,c_2,c_2)$& because $x_{n-3}\notin Ess(f_{n-2\leftarrow n-3})$\\
=$f(c_1,c_1,\ldots,c_1,c_1,c_1,c_1)$& because $x_{n-1}\notin Ess(f_{n \leftarrow n-1})$.
  \end{tabular}
  
 {\bf Case C.} Let $gap(f)=2$, $n$ be  odd and $ind(f)<\frac{n-1}{2}$.

 Let $ind(f)=\frac{n-m}{2}<\frac{n-1}{2}$, for some odd natural number  $m$, $n-2\geq m\geq 3$.
Let $h\in Min(f)$ be a minor of $f$ with $depth(h)=\frac{n-m}{2}$. Since $f$ is symmetric and  $gap(f)=2$ we have $x_v\notin Ess(f_{u\leftarrow v})$ for all $1\leq u,v\leq n$, $u\neq v$. Hence from Lemma \ref{l2.1} it follows that 
\\
\begin{tabular}{ll}
$h=$ & $[\ldots[f_{2\leftarrow 1}]_{4\leftarrow 3}\ldots]_{n-m\leftarrow n-m-1}=$ \\ &$f(x_1,x_1,x_3,x_3,\ldots,x_{n-m-1},x_{n-m-1},x_{n-m+1},\ldots,x_n)=$ \\ & $f(c_1,\ldots,c_1,x_{n-m-2},\ldots,x_n)$ 
\end{tabular}
\\
 for an arbitrary constant $c_1\in K$. Since $depth(h)=\frac{n-m}{2}$ and $m\leq n-2$ it follows that $Ess(h)=\emptyset$. Consequently,
$h=f(c_1,\ldots,c_1)=f(c_2,\ldots,c_2)$ for all $c_1,c_2\in K$.

$(ii)$ Let $n$, $3\leq n\leq k$ be an odd natural number, $gap(f)=2$ and $ind(f)=\frac{n-1}{2}$.

First, let  $n=3$. Then  from $(\ref{eq2.3})$ and $(\ref{eq2.4})$  it follows
that $f(i,i,i)=a_i$ and there are  $a_i,a_j$, $i,j\in K$ with $a_i\neq a_j$.  Hence   
$f(i,i,i)\neq f(j,j,j)$.

Second, let  $n>3$ and $ind(f)=\frac{n-1}{2}$. Let $g\in Min(f)$ be a minor of $f$ for which $depth(g)=ind(f)$ and as above we can write
\[g=[\ldots[f_{2\leftarrow 1}]_{4\leftarrow 3}\ldots]_{n-1\leftarrow n-2}.\]
  Let $h$ be a minor of $f$ with $depth(h)=\frac{n-3}{2}<ind(f)$  such that $g=h_{n-1\leftarrow n-2}$, i.e. $x_{n-2},x_{n-1}\in Ess(h)$ and by the symmetry of $f$ we have $\{x_{n-2},x_{n-1},x_n\}= Ess(h)$.

Then there is a ternary function $t\in P_k^3$  such that
\[t(x_{n-2},x_{n-1},x_{n})= h(a_1,\ldots,a_{n-3},x_{n-2},x_{n-1},x_{n})\] for
all $(a_1,\ldots,a_{n-3})\in K^{n-3}$ and $t$ is symmetric (see Remark
\ref{r2.1}).

Thus we have
$t(x_{n-2},x_{n-1},x_{n})=f(c_1,c_1,\ldots c_1,c_1,x_{n-2},x_{n-1},x_{n})$ for an arbitrary $c_1\in K$. Hence  $f(c,\ldots,c)=t(c,c,c)$ for all $c\in K$.
 If $x_u,x_v\in Ess(h)$ then $x_v\notin Ess(h_{u\leftarrow v})$,  else  $x_v\in Ess(f_{u\leftarrow v})$ which is impossible, according to Lemma \ref{l2.3}. If we suppose that $Ess(h_{u\leftarrow v})=\emptyset$, then by the symmetry of $f$ it follows that $depth(h)=ind(f)$ which is a contradiction. Again, by the symmetry of $f$ it follows that $Ess(h_{u\leftarrow v})=Ess(t_{u\leftarrow v})=Ess(t)\setminus\{x_u,x_v\}$ and hence $t\in G_{2,k}^3$. According to Theorem \ref{t2.3} it follows that there exist $c_1,c_2\in K$ such that $t(c_1,c_1,c_1)\neq t(c_2,c_2,c_2)$ (see case $n=3, gap(f)=2$) and hence  $f(c_1,\ldots,c_1)\neq f(c_2,\ldots,c_2)$.
\end{proof}
 \begin{theorem}\label{t2.5} Let  $3<min(n, k)$. If
  $f\in G_{2,k}^n$ is a symmetric function then 
\[f=\bigoplus_{i=1}^{n-1}\bigoplus_{j=i+1}^{n}\bigoplus_{m=0}^{k-1}x_i^mx_j^m
g(x_1,\ldots,x_{i-1},x_{i+1},\ldots,x_{j-1},x_{j+1},\ldots,x_n)\ \oplus\]
\[\oplus\ h(x_1,\ldots,x_n),\]
where $g$ and $h$ are symmetric functions such that: $h(\alpha)=0$ for all $\alpha\in Eq_k^n$ and 
\[g\in\left\{\begin{array}{ccc}
             G_{2,k}^{n-2}\  &\  if \  &\ ind(f)>2 \\
            G_{n-2,k}^{n-2}\  &\  if \  &\ ind(f)=2.  
                       \end{array}
           \right.\]

 \end{theorem}
\begin{proof}~~
 The conjunctions in $SC$-form of any function $f\in P_k^n$ can be reordered such that 
\[f=\bigoplus_{i=1}^{n-1}\bigoplus_{j=i+1}^{n}\bigoplus_{m=0}^{k-1}x_i^mx_j^m
g_{ijm}\ \oplus\ h(x_1,\ldots,x_n),\]
 $var(g_{ijm})=\{x_1,\ldots,x_{i-1},x_{i+1},\ldots,x_{j-1},x_{j+1},\ldots,x_n\}$
and  $h(\alpha)=0$
for all $\alpha\in Eq_k^n$.

Let $f\in G_{2,k}^n$ be a symmetric function with $n>2$.  Since   $h$ might assume non-zero values on the set  $Dis_k^n$, only, it follows that $h$ has to be a symmetric function.

 Then we obtain  
\[ f_{2\leftarrow
1}=\big[\bigoplus_{m=0}^{k-1}x_1^mx_1^m g_{12m}\big]\ \oplus\
\big[\bigoplus_{i=3}^{n-1}\bigoplus_{j=i+1}^{n}\bigoplus_{m=0}^{k-1}x_i^mx_j^m
[g_{ijm}]_{2\leftarrow 1}\big] \oplus\]
\[\oplus\ \bigoplus_{i=3}^{n}\bigoplus_{m=0}^{k-1}x_i^m
g_{1im}(x_2=m)\ \oplus \
\bigoplus_{i=3}^{n}\bigoplus_{m=0}^{k-1}x_i^m
g_{2im}(x_1=m).\]

 Since $x_v\notin Ess(f_{u\leftarrow v})$ for  $1\leq u,v\leq n$, $u\neq v$ it follows that $g_{12m}=g_{12s}$ for all $s,m\in K$. By symmetry of $f$ it follows that  $g_{ijm}=g_{ijs}$ for all $s,m\in K$ and $1\leq i<j\leq n$. Hence  the index $m$ is redundant and we might write $g_{ij}$ instead of $g_{ijm}$, i.e.  
$g_{ij}:=g_{ijm}$ for $m\in K$. The  symmetry of $f$ implies
$g_{ij}(\alpha)=g_{uv}(\alpha)$ for each $\alpha\in
K^{n-2}$, i.e. the functions $g_{ij}$ are identical, considered as mappings of $
K^{n-2}$ to $K$. Hence there is an $(n-2)$-ary function $g\in P_k^{n-2}$ which
maps each  $\alpha\in K^{n-2}$ as follows 
$g(\alpha)=g_{ij}(\alpha)$. 
Consequently  
 $g_{ij}=g(x_1,x_2\ldots,x_{i-1},x_{i+1},\ldots,x_{j-1},x_{j+1},\ldots,x_n)$    for  $ 1\leq i<j\leq n$.

Suppose that $g$ is not a  symmetric function. Without loss of generality assume
that $g_{ij}$ is not symmetric with respect to $x_1,x_2$ and $3\leq i<j\leq n$. 
Then there exist $n-2$ constants $c_1,c_2,c_3,\ldots,c_{n-2}\in K$ such that $g_{ij}(c_1,c_2,c_3,\ldots,c_{n-2})\neq
g_{ij}(c_2,c_1,c_3,\ldots,c_{n-2})$. Clearly $c_1\neq c_2$. If $d_1,d_2\in K$ with $d_1\neq d_2$ then 
\[f(x_1=d_1,x_2=d_2)=\bigoplus_{i=3}^{n-1}\bigoplus_{j=i+1}^{n}\bigoplus_{m=0}^{
k-1}x_i^mx_j^m g_{ij}(x_1=d_1,x_2=d_2)\ \oplus \] \[\oplus\
h(x_1=d_1,x_2=d_2).\] Since  $h$ is
symmetric,  it follows $h(x_1=d_1,x_2=d_2)=h(x_1=d_2,x_2=d_1)$ and hence 
$f(x_1=c_1,x_2=c_2)\neq f(x_1=c_2,x_2=c_1)$ which is a contradiction.

 Hence $g_{ij}$ is a symmetric
$(n-2)$-ary function which essentially depends on all of its variables.
Since $ess(f_{2\leftarrow 1})=n-2$ it follows that $x_1\notin Ess([g_{ij}]_{2\leftarrow 1})$ and hence $gap(g_{ij})>1$. According to Lemma \ref{l2.2} we have $gap(g_{ij})=2$ or $gap(g_{ij})=n-2$.

Let $ind(f)>2$. Then $ess([f_{2\leftarrow 1}]_{4\leftarrow 3})>0$ implies $Ess([g_{ij}]_{2\leftarrow 1})\neq\emptyset$. By the symmetry of $f$ and $g_{ij}$ it follows that  $Ess([g_{ij}]_{2\leftarrow 1})=\{x_3,\ldots,x_{n}\}\setminus\{x_i,x_j\}$. Hence $g_{ij}\in G_{2,k}^{n-2}$ for  $ 1\leq i<j\leq n$.

Let $ind(f)=2$. Then $ess([f_{2\leftarrow 1}]_{4\leftarrow 3})=0$ implies $Ess([g_{ij}]_{2\leftarrow 1})=\emptyset$.  Hence $g_{ij}\in G_{n-2,k}^{n-2}$ for  $ 1\leq i<j\leq n$. 

\end{proof}

 Theorem \ref{t2.2},  Theorem \ref{t2.3} and  Theorem \ref{t2.5} provide
decompositions of the symmetric functions with non-trivial arity gap 
 in the basis $\langle\oplus,\cdot\  ,\{x^\alpha\}_{\alpha\in K}
\rangle$ of the algebra $P_k^n$.

As usual we shall say that a $k-$valued function $f\in P_k^n$ is {\it linear}
if 
$f=a_1x_1\oplus a_2x_2\oplus\ldots\oplus a_nx_n\oplus c,$
where $a_1,a_2,\ldots a_n,c\in K$.
Clearly, $x_i\in Ess(f)$ if and only if $a_i\neq 0$.

 \begin{theorem}\label{t2.6}
 The set  $P_{k}^n$, $k,n\geq 2$, contains a linear  function with non-trivial arity gap if
and only
if $k$ is an even natural number.  
 \end{theorem}
\begin{proof}
  Let $f=\big[\bigoplus_{i=1}^na_ix_i\big]\ \oplus c$ with $c,a_i\in K$. Without loss of generality let us consider the identification minor $f_{2\leftarrow 1}= (a_1\oplus a_2)x_1\ \oplus\ \big[\bigoplus_{i=3}^na_ix_i\big]\ \oplus c$. Clearly $ess(f_{2\leftarrow 1})\geq ess(f)-2$, i.e. $gap(f)\leq 2$. 

 Let $k$ be an even natural number and $k=2m$ for some $m\in N$. Then let us
consider the following linear (and symmetric) function
\\ \centerline{$f=m(x_1\oplus x_2\oplus\ldots\oplus x_n)\oplus c,$}
for some $c\in K$.
Clearly,
\\ \centerline{$f_{i\leftarrow j}=m(x_1\oplus\ldots\oplus
x_{j-1}\oplus x_{j+1}\oplus\ldots x_{i-1}\oplus x_{i+1}\oplus\ldots \oplus
x_n)\oplus c.$}  Hence   $f\in G_{2,k}^n$.

 Let     $k$ be an odd natural number and let $f=a_1x_1\oplus\ldots\oplus
a_nx_n\oplus c$, for some $c\in K$, be a linear $k-$valued function. First assume that there are
$i$ and $j$, $1\leq i,j\leq n$, such that $i\neq j$ and $a_i=a_j\neq 0$.
Without loss of generality let us assume $(j,i)=(1,2)$. Then we
have $a_1\oplus a_2=2a_1$ and 
\\ \centerline{$f_{2\leftarrow 1}=2a_1x_1\oplus a_3x_3\oplus\ldots\oplus
a_nx_n\oplus c.$}
 Since $k$ is odd it follows that $2a_1\neq 0\ (mod\ k)$. Hence $Ess(f_{2\leftarrow
1})=\{x_1,\ldots,x_n\}\setminus\{x_2\}$ and  $f\notin G_{2,k}^n$.
Second, let   $a_i\neq a_j$ for all  $i$ and $j$, $1\leq i<j\leq n$.
Then we have $a_1+a_2\neq k$ or $a_1+a_3\neq k$. Without loss of generality
assume that $a_1+a_2\neq k$. Hence
\\ \centerline{$f_{2\leftarrow 1}=(a_1+a_2)x_1\oplus a_3x_3\oplus\ldots\oplus
x_n\oplus c.$} Since $k\neq a_1+a_2<2k$ it follows that $a_1+a_2\neq 0\ (mod\
k)$
which implies $f\notin G_{2,k}^n$.
\end{proof}

One can prove that if $f$ is a linear function with non-trivial arity gap then $f$ is symmetric.

\section{Subfunctions of  symmetric functions  with non-trivial arity gap}\label{sec3}

In this section, we shall study the subfunctions of the symmetric $k$-valued
functions $f$ with non-trivial arity gap.
 
Let $c\in K$ be a constant from $K$ and $f\in P_k^n$ be a symmetric function. We
say that $c$ is a
\emph{ dominant} of $f$ if
$f(c_1,\ldots,c_{n-1},c)=f(d_1,\ldots,d_{n-1},c)$ for every
 $c_1,\ldots,c_{n-1},d_1,\ldots,d_{n-1}\in K$. $Dom(f)$ denotes the set of all dominants of $f$.
 
 Clearly if $c\in Dom(f)$ then
$Ess(f(x_1,\ldots,x_{n-1},c))=\emptyset$, i.e. the subfunctions of $f$ of order $1$ obtained by dominants of $f$ are always constant functions.  If $f\in G_{n,k}^n$ then  $c\in Dom(f)$ if and only if  $f(c_1,\ldots,c_{n-1},c)=f(0,\ldots,0)$ for all $c_1,\ldots,c_{n-1}\in K$,
according to Theorem \ref{t2.1}.

A constant $c\in K$  is called \emph{weak dominant} of $f$ if it is a dominant of an identification minor of $f$. 

If $f$ is a symmetric function then weak dominants of $f$ are dominants of all identification minors of $f$. $Wdom(f)$ denotes the set of all weak dominants of $f$.

\begin{theorem}\label{t3.1}
Let $f\in G_{n,k}^n$ be a  symmetric function with $2\leq k$, $2<n$ and let $g=f(x_i=c)$ for some $x_i$,  $1\leq i\leq n$ and for some constant $c\in K$ be a subfunction of $f$.  If $c\notin Dom(f)$ then $g$ is  a symmetric function which belongs to the class $G_{n-1,k}^{n-1}$.
\end{theorem}
\begin{proof} We shall consider the non-trivial case $n>2$ (else the subfunctions of $f$ will depend on at most one essential variable). Hence $k>2$ because $2<n=gap(f)\leq k$.

By symmetry we may assume that $g=f(x_n=c)$. Since $c\notin Dom(f)$ it follows $Ess(g)\neq\emptyset$. Lemma \ref{l2.1} implies that $Ess(g)=\{x_1,\ldots,x_{n-1}\}$.
Thus we obtain
\[g_{2\leftarrow
1}=g(x_1,x_1,x_3,\ldots,x_{n-1})=f(x_1,x_1,x_3,\ldots,x_{n-1},c).\] Theorem
\ref{t2.2} implies that for every $n-2$ constants
$c_1,\ldots,c_{n-2}\in K$ we have 
\[g(c_1,c_1,c_2,\ldots,c_{n-2})=f(c_1,c_1,c_2
,\ldots,c_{n-2},c)= f(0,\ldots,0)\] because
$(c_1,c_1,c_2,\ldots,c_{n-2},c) \in Eq_k^n$. Consequently 
$g$ is symmetric and $g\in G_{n-1,k}^{n-1}$.
 
\end{proof}

 Let us denote $range(f)=|\{f(\alpha)\ |\ \alpha\in K^n\}|$ for $f\in P_k^n$ and 
\[sub_k^n=\binom{k}{1}+\binom{k}{2}+\ldots+\binom{k}{n-1}.\]

\begin{lemma}\label{l3.1}
If $f\in G_{n,k}^n$, $n\leq k$ is a symmetric function, then 
$sub(f)\leq sub_k^n+range(f)$.
\end{lemma}
\begin{proof} Let $f\gg g$ and $ord(g)=m>1$. Without loss of generality let us assume 
 $g=f(x_1=c_1,\ldots,x_m=c_m)$. 

Let $\alpha=(c_1,\ldots,c_m)\in Eq_k^m$. Then Theorem
\ref{t2.2} implies that $g=f(0,\ldots,0)$, i.e. $g$ is a constant.
So, $g$ can be obtained in two ways, only: when
$\alpha\in Eq_k^m$ or $m=n$. Then it is clear that the number of all constant
subfunctions is equal to $range(f)$.

Let $\alpha=(c_1,\ldots,c_m)\in Dis_k^m$. Since
$|Dis_k^m|=\binom{k}{m}.m!$,  the symmetry of
$f$
implies that there exist at most
$\binom{k}{m}$ subfunctions of
order $m$, $1\leq m\leq n-1$. 
Thus, 
if $f\in G_{n,k}^n$, $n\leq k$ is a symmetric function then the number of all its subfunctions is equal to  at most $sub_k^n$. Hence $sub(f)\leq sub_k^n+range(f)$.
\end{proof}
\begin{remark}\label{r3.1}\rm ~~~~

$(i)$ Note that Lemma \ref{l2.1} and Theorem \ref{t3.1} imply that if $g\in Sub(f)$ with $ess(g)=l>1$ then $g\in G_{l,k}^l$.
 
$(ii)$  Let $f$ be a function represented as in (\ref{eq2.2}) with $a_0=0$ and let $b_{\beta}\in K$ be non-zero integers for all $\beta\in Dis_k^n$.
 Let $(c_{m+1},\ldots,c_n)\in Dis_k^{n-m}$ and $m<n$. Then we have 
$f(x_{m+1}=c_{m+1},\ldots,x_n=c_n)=\bigoplus_{\gamma\in Dis_k^{m}}b_{\alpha}S(m,{\gamma}),$
where ${\alpha}=(d_1,\ldots,d_m,c_{m+1},\ldots,c_{n})\in Dis_k^n$ and $\gamma=(d_1,\ldots,d_{m})\in Dis_k^{m}$. Since $b_{\beta}\neq 0$, it follows that $f(x_{m+1}=c_{m+1},\ldots,x_n=c_n)$ depends essentially  on all its $m$ variables.  Consequently, $f(x_{m+1}=c_{m+1},\ldots,x_n=c_n)=f(x_{m+1}=a_{m+1},\ldots,x_n=a_n)$ for $a_{m+1},\ldots,a_n\in K$ if and only if $\{c_{m+1},\ldots,c_n\}=\{a_{m+1},\ldots,a_n\}$. This implies that $sub(f)=sub_k^n+range(f)$, i.e. the function $f$ reach the upper bound for $sub(f)$, obtained in Lemma \ref{l3.1}.

$(iii)$ The next example will show that $sub(f)< sub_k^n+range(f)$ can be happen. Let $k=4$, $n=3$ and $f=S(3,(0,1,3))\oplus S(3,(0,2,3)) (mod\ 4)$.
Clearly, $f\in G_{3,4}^3$ and $f(x_1=0,x_2=1)=f(x_1=0,x_2=2)=x_3^3,$
$f(x_1=1,x_2=3)=f(x_1=2,x_2=3)=x_3^0,$ and $f(x_1=1,x_2=2)=0,$ which shows that
$sub(f)=4+3+3=10$ and $sub_4^3=\binom{4}{1}+\binom{4}{2}+range(f)=4+6+3=13$. 
 
\end{remark}

\begin{theorem}\label{t3.2}
 Let $f\in G_{2,k}^n$, $3\leq min(n,k)$ be a symmetric function and $c\in K$. Then
 
$(i)$ $t=f(x_{i}=c,x_{j}=c)\in G_{2,k}^{n-2}$ for all $i,j$, $1\leq i,j\leq n$, $i\neq j$ if $ind(f)>2$;

$(ii)$  $t=f(x_{i}=c,x_{j}=c)\in G_{n-2,k}^{n-2}$ for all $i,j$, $1\leq i,j\leq n$, $i\neq j$ if $ind(f)=2$; 

$(iii)$ $t=f(x_{i}=c)\in G_{n-1,k}^{n-1}$ for all $i$, $1\leq i\leq n$ if $c\in Wdom(f)$; 

$(iv)$  $t=f(x_{i}=c)\in G_{2,k}^{n-1}$ for all $i$, $1\leq i\leq n$ if $c\notin Wdom(f)$.

 \end{theorem}
\begin{proof}
Let $f\in G_{2,k}^n$, $3\leq min(n,k)$ be a symmetric function and $c\in K$. By symmetry of $f$ we might consider the pair $(1,2)$ instead $(i,j)$.

 $(i)$ From  Lemma \ref{l2.5} and   $ind(f)>2$  it follows that $n\geq 6$. Then
we have    $t(a_1,\ldots,a_{n-2})=h(c,c,a_1,\ldots,a_{n-2})$ for all
$(a_1,\ldots,a_n-2)\in K^{n-2}$, where $h=f_{2\leftarrow 1}$ and
$depth(h)=1<ind(f)$. From Lemma \ref{l2.3} it follows that $t$ and $h$ depends
on $n-2$ variables, i.e. $Ess(h)=\{x_3,\ldots,x_n\}$. Then $g=h_{4\leftarrow
3}=[f_{2\leftarrow 1}]_{4\leftarrow 3}$ is a minor of $f$ with
$depth(g)=2<ind(f)$. Hence it follows that $Ess(g)\neq \emptyset$ and by
symmetry of $f$ we have $Ess(g)=Ess(t_{4\leftarrow 3})=\{x_5,\ldots,x_n\}$.
Hence $t\in G_{2,k}^{n-2}$.

$(ii)$ Let $ind(f)=2$ and  $t$, $h$ and $g$ are as in $(i)$.  Now, $depth(g)=2=ind(f)$  implies that $Ess(g)=Ess(t_{4\leftarrow 3})=\emptyset$. By symmetry of $f$ it follows that all identification minors of $t$ do not depend on any of its variables. Hence $t\in G_{n-2,k}^{n-2}$. 

$(iii)$ Let $c\in Wdom(f)$ and $t=f(c,x_2,\ldots,x_n)$. Without loss of generality,  assume that $c$ is a dominant of $f_{n\leftarrow n-1}$, i.e. $f(c,x_2,\ldots,x_{n-2},c_1,c_1)$ does not depend essentially on any variable for all $c_1\in K$. Then  Lemma \ref{l2.3}  implies $t_{3\leftarrow 2}=f(c,c,c,x_4,\ldots,x_n)=f(c,x_2,\ldots,x_{n-2},c_1,c_1)$. Hence  $Ess(t_{3\leftarrow 2})=\emptyset$, i.e.
$t\in G_{n-1,k}^{n-1}$. 

$(iv)$ Let $c\in K$ and $c\notin Wdom(f)$ and $t=f(c,x_2,\ldots,x_n)$. Then $t_{3\leftarrow 2}=f(c,c,c,x_4,\ldots,x_n)$ depends on at least one variable (else $c\in Wdom(f)$). From Lemma \ref{l2.1} it follows that $Ess(t_{3\leftarrow 2})=\{x_4,\ldots,x_n\}$ and hence $t\in G_{2,k}^{n-1}$.
\end{proof}

\begin{corollary}\label{c3.1}
 If $f\in P_{k}^n$,   is a symmetric function with non-trivial arity gap, then each its subfunction $g=f(x_n=c)$ with $c\notin Dom(f)$  has non-trivial arity gap.
\end{corollary}
\begin{proof}
 If $f\in G_{n,k}^n$ we are done by Theorem \ref{t3.1} and if $f\in G_{2,k}^n$ by Theorem \ref{t3.2}.
\end{proof}

\section{Separable sets of symmetric functions with non-trivial arity gap}\label{sec4}

\begin{definition}\label{d4.1}
A set $M$ of essential variables in $f$ is called {\it separable in $f$} if
there is a subfunction $g$ of $f$ such that $M=Ess(g)$.

$Sep(f)$ denotes the set of all separable sets in $f$ and $sep(f)=|Sep(f)|$.
\end{definition}
Note that the constants in the range $V(f)=\{c\in K\ |\ \exists \alpha\in K^n, \ \ f(\alpha)=c\}$ of $f$   form subfunctions of $f$, which do not depend on any essential variable. So, the empty set, we will include in $Sep(f)$.

The numbers $sep(f)$ and $sub(f)$ are important complexity measures of a
function $f\in P_k^n$. The separable sets and the valuations $sep(f)$ and $sub(f)$ are studied in work of many authors  O. Lupanov \cite{lup},  K. Chimev \cite{ch3,ch51}, A.
Salomaa \cite{sal}, S. Shtrakov and K. Denecke
\cite{sh51}, etc. 

If $f\gg g$ with $ord(g)=m>0$ then $g$ uniquely determines an $m-$element set $M$,  $M=Ess(g)\subseteq
Ess(f)$, which is separable in $f$. It is possible  the same $M$ to be the set of essential
variables of another subfunction $t$, $f\gg t$ of $f$, i.e. $Ess(g)=Ess(t)$,
but $g\neq t$. Consequently, $sep(f)\leq sub(f)$. Theorem \ref{t3.1} and
Theorem \ref{t3.2} show that if $f$ is a symmetric function with non-trivial
arity gap then its subfunctions with respect constants outside $Dom(f)$ have non-trivial arity gap. Lemma \ref{l3.1} gives an upper bound of $sub(f)$.

In this section, we prove that the complexity measure $sep(f)$ assumes its
maximum value on the symmetric functions with non-trivial arity gap.
\begin{theorem}\label{t4.1}
If $f$ is a symmetric function with non-trivial arity gap, then each set of essential variables in $f$ is separable in $f$.
\end{theorem}
\begin{proof} Let $f\in G_{n,k}^n$, $n\leq k$ and 
let $Ess(f)=\{x_1,\ldots,x_n\}$. Without loss of generality let us prove
that $M=\{x_1,\ldots,x_m\}$, $m<n$ is a separable set in $f$. According to
(\ref{eq2.1}) there are  constants $c_1,\ldots,c_n\in K$ such that
$f(c_1,\ldots,c_n)\neq a_0,$ where $a_0=f(d_1,\ldots,d_n)$ for all
$(d_1,\ldots,d_n)\in Eq_k^n$.
We have to show that if $f_1:=f(x_{m+1}=c_{m+1},\ldots,x_n=c_n)$ then $M=Ess(f_1)$.
Let $x_t\in M$ be an arbitrary variable from $M$, i.e. $1\leq t\leq m$. Again
from (\ref{eq2.1}) it follows that 
\\
\centerline{$ f(c_1,\ldots,c_{t-1},c_n,c_{t+1},\ldots,c_m,\ldots,c_n)=a_0.$}
Hence $x_t\in Ess(f_1)$ which implies $M=Ess(f_1)$.

Let $f\in G_{2,k}^{n}$, $n\leq k$ be a symmetric function.
 Without loss of generality let us assume that $M=\{x_1,\ldots,x_m\}$, $m<n$ is a set of essential variables in $f$. We have to prove that $M$ is a separable set in $f$. Since $x_1\in Ess(f)$ by Theorem 1.2  \cite{ch51}, there is a chain of subfunctions
\[f=f_n\rhd f_{n-1}\ldots\rhd f_2\rhd f_1\]
such that $Ess(f_1)=\{x_1\}$ and $Ess(f_j)=\{x_1,x_{i_2},\ldots,x_{i_j}\}$ for all $j=2,3,\ldots,n$. Without loss of generality we might assume that $i_l=l$ for $l=2,\ldots,j$ and  there are constants $c_{m+1},\ldots,c_n$ for the variables in $Ess(f)\setminus Ess(f_m)$ such that 

\centerline{$f_m=f(x_{{m+1}}=c_{m+1},x_{{m+2}}=c_{m+2},\ldots,x_{{n}}=c_{n}).$}
Consequently, $f(x_{{m+1}}=c_{m+1},x_{{m+2}}=c_{m+2},\ldots,x_{{n}}=c_{n})$ is a
function which depends essentially on the variables $x_1,\ldots,x_m$, i.e. $M$
is a separable set in $f$.
\end{proof}
\begin{corollary}\label{c4.1}
 If $f\in P_k^n$ is a symmetric function with non-trivial arity gap then $sep(f)=2^n$.
\end{corollary}
\begin{corollary}\label{c4.2}
 If $f\in P_k^n$ is a symmetric function with non-trivial arity gap  then $sub(f)\geq 2^n$.
\end{corollary}
Lemma \ref{l3.1} implies that if $n\leq k$ and $f\in G_{n,k}^n$ then 
\[2^n=sep(f)\leq sub(f)\leq \sum_{i=1}^{n}\binom{k}{i}\] and if $k=n$ then
$2^n=sep(f)= sub(f)$.

\end{document}